# Detection of Thalassaemia Carriers by Automated Feature Extraction of Dried Blood Drops

*Manikuntala Mukhopadhyay[1], Manish Ayushmann[1], Pourush Sood[2], Rudra Ray[3], Maitreyee Bhattacharyya[3], Debasish Sarkar[4] and Sunando Dasgupta[1*]*

1. Department Of Chemical Engineering, Indian Institute Of Technology, Kharagpur, Pin 721302, West Bengal, India.
2. Department of Electronics and Electrical Communication Engineering, Indian Institute Of Technology, Kharagpur, Pin 721302, West Bengal, India.
3. Institute of Hematology & Transfusion Medicine, Medical College, Kolkata, Pin 700073, West Bengal, India.
4. Department of Chemical Engineering, University of Calcutta, Pin 700009, West Bengal, India.

*Corresponding author

Email: sunando@che.iitkgp.ac.in

Ph: + 91-3222-283922

**Abstract**

Thalassaemia, triggered by defects in the globin genes, is one of the most common monogenic diseases. The β-thalassaemia carrier state is clinically asymptomatic, thus, making it onerous to diagnose. The current gold standard technique is implausible to be used for onsite carrier detection as the method necessitates expensive instruments, skilled manpower and time. In this study, we have tried to classify the carriers from the healthy samples based on their blood droplet drying patterns using image analysis based tools and subsequently develop an in-house program for automated classification of the same. This automatic, rapid, less laborious and cost-effective technique will significantly increase the total number of carriers that are screened for thalassaemia per year in the country, thus, reducing the burden in the state run advanced health facilities.

# 1. Introduction

Thalassaemia is one of the most common single gene disorders in which the normal production of haemoglobin is impaired. This autosomal recessive disorder decreases the synthesis of either the α- globin or β- globin chains that compose adult haemoglobin, HbA ($\alpha_2\beta_2$)[1], thus causing imbalance in the globin chain synthesis. This syndrome is mostly prevalent in the Mediterranean, Middle East, tropical Africa, Indian subcontinent and Central Asia, and in aggregate is among the most common inherited disorders of humans. Indian populations account to ~10% of the world's thalassaemia carriers.

β- thalassaemia is caused by the inadequate synthesis ($\beta^+$) or absence ($\beta^0$) of the β globin chains[2] whereas deficient synthesis of α chains leads to α thalassaemia.[3] The hematological consequences of reduced synthesis of one globin chain results from haemoglobin deficiency and also from a relative excess of the other globin chain, especially in β-thalassaemia. The β-thalassaemia carrier state is clinically asymptomatic and thus largely remains undiagnosed. The major challenge in approaching 'zero-thalassaemia' nation is to screen the entire population under the reproductive age group. The gold standard method that is used for carrier detection in hospitals, at present, is High Performance Liquid Chromatography (HPLC) which is a laboratory based method requiring expensive instruments, skilled manpower and time, thus, making it difficult to be used as an onsite method. Also, Nestroft[4] can be used in some cases for detection of β-thalassaemia carriers. The inaccurate eye estimation, long standing hours and higher percentage of false negative results, however, made this technique inapposite for onsite detection of carriers. A rapid, portable and automated technology for thalassaemia carrier screening is hence of significant importance.

Detection of diseases from drying pattern of body fluids has recently been emerging as a tool for automated and rapid diagnosis of several diseases.[5,6,7] Whole blood, a complex colloidal suspension, behaves like a non-Newtonian fluid [8] and leads to the formation of several interesting and complex patterns when dried on a solid substrate. Numerous intra-cellular and macromolecular interactions in the drying droplets of blood or plasma stem up the final morphologies of the resulting dried patterns. The onset of any pathological abnormality alters the cellular and the macromolecular components of the blood, thus altering their drying patterns. This technique can hence be used as a simple diagnostic tool for the detection of several diseases based on the dried drop patterns of the healthy and the diseased subjects.

Specific diseases give rise to distinct drying patterns where the final patterns depend on several factors such as plasma content and the morphology of red blood cells. This technique was first developed in Russia with the 'Litos' test system[9] where urine samples of patients suffering from kidney stones were found to leave a distinct drying pattern.[5, 10] The efficacy of droplet drying of body fluids has also been studied for the detection of various diseases like hepatitis, breast cancer, leukemia, lung cancer, pathogen infection, premature delivery and kidney diseases.[5,7,11,12] Recently the potency of the technique to diagnose diseases based on the whole blood drop drying patterns was extensively investigated by Brutin et al.[11] where it was found that the dried patterns of a drop of blood from a healthy subject was markedly different from that of a patient suffering from anaemia and hyperlipidaemia. Yakhno *et al.* carried out several investigations [9,10,12,13] studying the effect of various factors influencing the drying pattern of biological fluids. In one of his recent investigations, patterns observed by drying droplets of biological fluids from cattle were used for the detection of leukemia and tuberculosis[14].

There is an urgent need in the developing countries to revolutionize the healthcare industry with simple, cost effective and non-invasive techniques for disease diagnosis. In this aspect, several researchers have come up with many innovative lab-on-chip devices where a portable paper based device or a smartphone can rapidly detect the serum cholesterol, perform genetic testing and can detect malaria even in limited resource settings.[15-18] With the recent advancements in the field of microfluidics, a simple point of care device can now diagnose and monitor jaundice, detect analytes in blood and can even identify sickle cell disease based on the density differences of the red blood cells [19-21]. Till date no attempt has been made to detect the thalassaemia carriers by exploiting the blood droplet drying technique which being both rapid & less laborious can potentially aid in the screening of the carrier samples of the entire population under reproductive age. In this study, the aim is to separate the thalassaemia carriers from the healthy subjects based on the blood droplet drying patterns using image analysis based tools to identify the distinct signatory features in the patterns of the respective samples. The proposed technique can be instrumental in developing a rapid, less laborious and cost effective 'on-field' screening method which will enable the initial screening of carriers from the large population.

## 2. MATERIALS AND METHODS

**2.1 Sample Collection:** Thalassaemia carrier samples were collected from the individuals, family members of the thalassaemia patients visiting Institute of Haematology and Transfusion medicine (IHTM), Medical College, Kolkata for thalassaemia screening and also from outreach camps, conducted by IHTM routinely. The study population consisted of a total of 100 carrier and healthy samples. Healthy samples were also collected from the camps conducted by IHTM. Informed consent and clinical history of the samples were recorded. Thorough clinical examinations for pallor, jaundice, and splenomegaly were done. 2ml of blood samples were collected in EDTA coated vials from the peripheral veins of individuals.

The individuals (non-smokers, age within 20 to 40 yrs, body weight 50±20 kg, resting heart rate being 73±4 beats/minute, systolic pressure-120.5±7 mm Hg and diastolic pressure being 84±3 mm Hg, devoid of any known disease & disorders) with normal ferritin value (> 30 ng/L) and detecting beta/HbE trait by HPLC analysis (HbA2 value 4.5 – 7.5, HbA2E value 20.0 – 30.0) were included in the group of carriers.

The individuals (non-smokers, age within 20 to 40 yrs, body weight 50±20 kg, resting heart rate being 73±4 beats/minute, systolic pressure-120.5±7 mm Hg and diastolic pressure being 84±3 mm Hg, devoid of any known disease & disorders) with normal HPLC parameters, ferritin value (> 30 ng/L), and normal RBC indices were considered as healthy.

**2.2 Haematological Analysis:** Complete hemogram was performed on Sysmex (KX- 21) automated cell counter. Haemoglobin analysis was estimated by high performance liquid chromatography (HPLC, from Bio Rad Variant II) and Feritin level was measured by ELISA based method. The Mean Corpuscular Haemoglobin (MCH) and the Mean Corpuscular Volume (MCV) of the healthy and the carrier samples have been reported in Table 1.

**2.3 Sample Characterization:** The pertinent physiochemical properties of the blood samples were evaluated. The first of these was the estimation of the surface tension by placing sessile droplets (2 µL) of both the healthy and carrier samples on cleaned glass substrates by using a Goniometer (*250 G1, Ram'ehart, Germany*). The numerical values of the surface tension are reported in Table I. The zeta potentials of the samples were evaluated (*Malvern Instruments, Model: ZEN 3600*) and were found to be negatively charged with the order of the surface potential ($\phi$, mV) being $\phi_{Healthy} > \phi_{Carrier}$.

Table 1: Characterization of the samples

| Sample Type | Surface Tension (mN/m) | Zeta Potential (mV) | MCV (fl) | MCH (pg) |
|---|---|---|---|---|
| Healthy | 52 | -12 | 90 | 30 |
| Carrier | 51 | -9 | 70 | 24 |

**2.4 Substrate preparation for droplet drying:**

Glass slides (Material Composition: $SiO_2$ - 81%, $B_2O_3$ - 13%, $Na_2O + K_2O$ – 4%, $Al_2O_3$ – 2%) procured from *HiMedia Laboratories Pvt. Ltd.* have been used as substrates. They were subjected to a thorough cleaning protocol that comprised of ultrasonication with acetone, followed by de-ionized water (10 minutes each). These cleaned slides were dried in a hot air oven at 95ºC to ensure complete removal of the surface moisture.

**2.5 Droplet drying & Imaging:** Sessile droplets (whole blood of both healthy and carrier samples) of 2 μl volume were allowed to dry on the cleaned glass substrates. The final dried patterns were then observed under the optical microscope (LEICA DMi8 A) in transmission mode with 5x magnification. All the experiments were performed at room temperature (~ 25ºC) and within 72 hours of the sample collection. The high quality images were analyzed in ImageJ software (v: 1.47 platform) to evaluate the average crack lengths of the dried blood drops.

## 3. RESULTS

**3.1 Droplet Drying Patters Of Whole Blood on Glass substrates**

Whole blood droplets of 2μl volume were allowed to dry naturally at room temperature with controlled humidity on the cleaned glass substrates. To check the reproducibility of the observed patterns, 6 drops of each blood sample were dried and analysed in each set. The average time required for the drops to dry on the glass slides was around 20-25 minutes. The study population consisted of 100 samples consisting of both healthy and carrier samples. It was observed that the radial cracks in the drying patterns of the healthy blood samples were significantly longer as compared to those in the carriers (Figures 1-4). The area of the central core was distinctly more in the carrier samples where more branching from the parent cracks were observed as compared to the patterns obtained from the healthy samples.

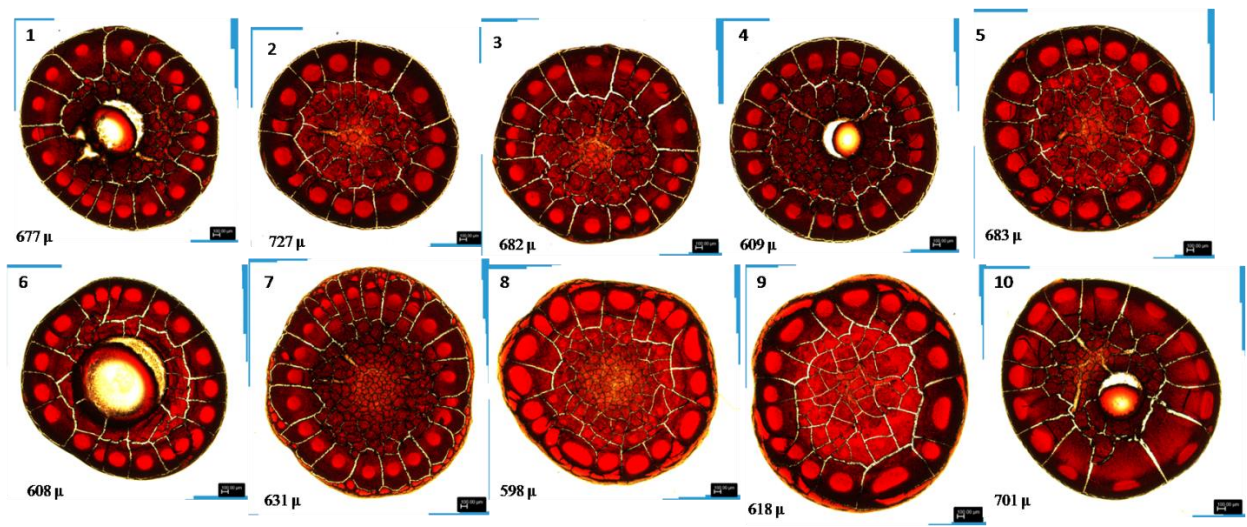

Figure 1: Dried blood drop patterns of carrier samples (the numbers on the top left denote the serial numbers whereas the ones below denote the average crack lengths)

(The dried drop blood patterns of the remaining carrier samples have been shown in Figure S1 of the ESI)

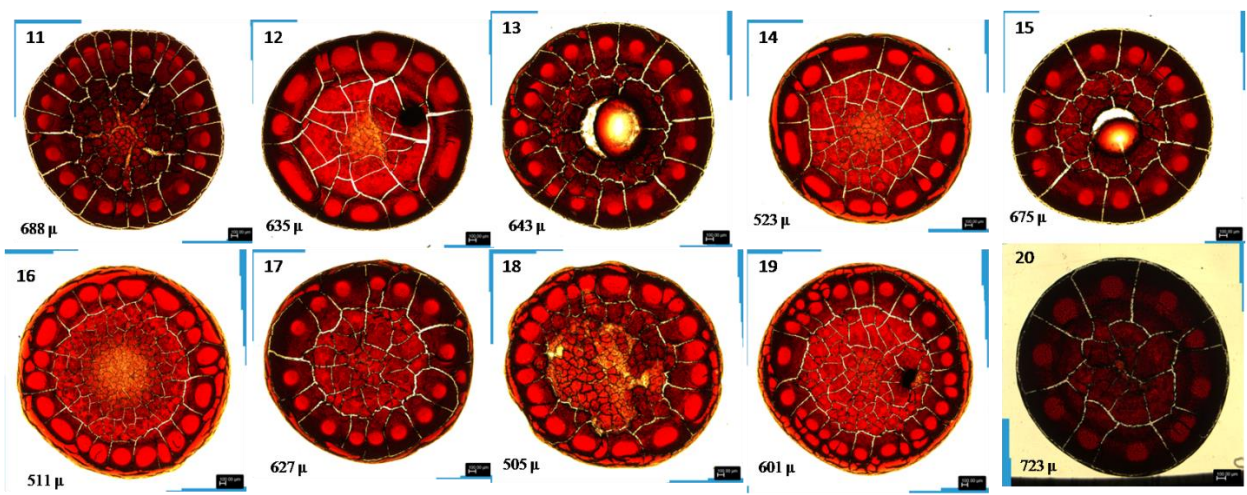

Figure 2: Dried blood drop patterns of carrier samples (the numbers on the top left denote the serial numbers and the ones below denote the average crack lengths)

(The dried drop blood patterns of the remaining carrier samples have been shown in Figure S1 of the ESI)

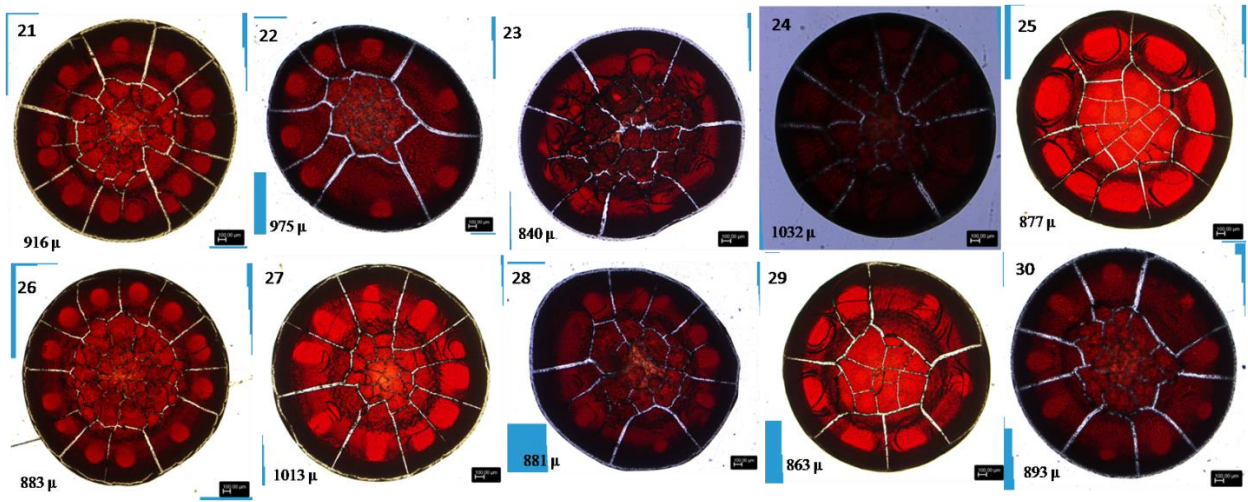

Figure 3: Dried blood drop patterns of healthy samples (the numbers on the top left denote the serial numbers and the ones below denote the average crack lengths)

(The dried drop blood patterns of the remaining carrier samples have been shown in Figure S2 of the ESI)

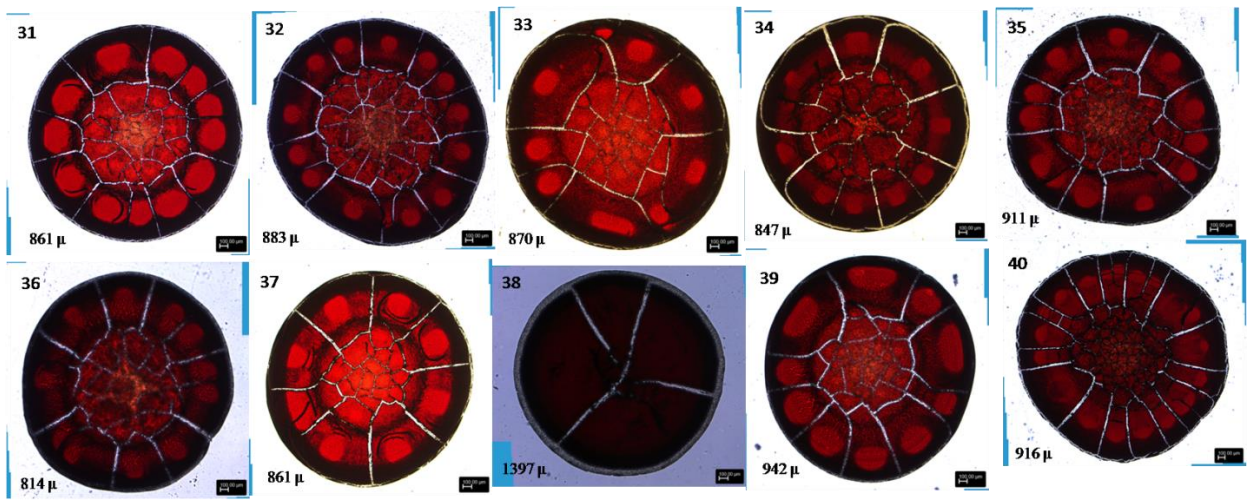

Figure 4: Dried blood drop patterns of healthy samples (the numbers on the top left denote the serial numbers and the ones below denote the average crack lengths)

(The dried drop blood patterns of the remaining carrier samples have been shown in Figure S2 of the ESI)

In order to ascertain the accuracy of this detection method based on the image analysis technique, further validation was done by comparing the results with the with the gold standard methods, as performed by clinicians at IHTM and have been produced below in Tables 2 and 3.

Table 2: Classification of carrier samples by both analyzing the crack lengths and the clinical diagnostic method (The analysis of the remaining carrier samples have been shown in table S3 of the ESI)

| SL.No | Age | Sex | **Diagnosis** | HPLC Test No. | HB | MCV | MCH | HBA0 | **HBA2** | HBF | **HBA2e** | **Crack length (in microns)** |
|---|---|---|---|---|---|---|---|---|---|---|---|---|
| 1 | 24 | Male | **HbE carrier** | 922 | 13.5 | 82.8 | 24.5 | 63 | | 1.1 | **26.3** | **677** |
| 2 | 27 | Female | **Beta thalassaemia carrier** | 937 | 11.1 | 83.2 | 25.2 | 80.5 | **5.3** | 2.5 | | **727** |
| 3 | 29 | Male | **Beta thalassaemia carrier** | 954 | 11.1 | 61.9 | 17.8 | 84.6 | **5.4** | 0.8 | | **682** |
| 4 | 32 | Male | **HbE carrier** | 966 | 14.1 | 89.1 | 28 | 63.1 | | 0.7 | **27.8** | **609** |
| 5 | 26 | Female | **HbE carrier** | 974 | 10.8 | 87 | 26.5 | 64.1 | | 1.3 | **26.6** | **683** |
| 6 | 22 | Female | **HbE carrier** | 983 | 11.7 | 81.5 | 24.3 | 65.3 | | 0.7 | **26.1** | **608** |
| 7 | 36 | Male | **Beta thalassaemia carrier** | 989 | 14.3 | 82.2 | 24.9 | 84.9 | **5.1** | 0.3 | | **631** |
| 8 | 25 | Female | **HbE carrier** | 996 | 9.4 | 90.4 | 27.2 | 67.3 | | 1 | **25.6** | **598** |
| 9 | 28 | Female | **HbE carrier** | 1017 | 10.1 | 76.3 | 25.4 | 64.4 | | 0.4 | **27.3** | **618** |
| 10 | 25 | Male | **Beta thalassaemia carrier** | 1018 | 12 | 63.1 | 19.7 | 83.3 | **5.8** | 0.3 | | **701** |
| 11 | 38 | Male | **HbE carrier** | 1035 | 15 | 82.4 | 27.2 | 67.5 | | 0.4 | **26.4** | **688** |
| 12 | 24 | Female | **HbE carrier** | 1038 | 9 | 79.4 | 25.7 | 65.3 | | 0.5 | **26.2** | **635** |
| 13 | 23 | Male | **HbE carrier** | 1048 | 14 | 75.1 | 23.5 | 66 | | 0.4 | **25.5** | **643** |
| 14 | 26 | Female | **Beta thalassaemia carrier** | 1053 | 9.3 | 65.6 | 20.5 | 86.3 | **4.5** | 1.1 | | **523** |
| 15 | 38 | Male | **Beta thalassaemia carrier** | 1054 | 12 | 66.8 | 20.8 | 84.8 | **4.7** | 1.3 | | **675** |
| 16 | 24 | Female | **Beta thalassaemia carrier** | 1055 | 9.4 | 70.1 | 22.3 | 85.5 | **5.6** | 0.9 | | **511** |
| 17 | 19 | Female | **HbE carrier** | 1061 | 13 | 69.1 | 23.1 | 63.6 | | 0.6 | **28.3** | **627** |
| 18 | 20 | Female | **HbE carrier** | 1091 | 10 | 77.1 | 25.7 | 65 | | 0.9 | **26.3** | **505** |
| 19 | 25 | Female | **HbE carrier** | 1093 | 11 | 74.2 | 24.3 | 63.3 | | 1.1 | **28.5** | **601** |
| 20 | 20 | Female | **HbE carrier** | 303 | 10.9 | 76.4 | 24.3 | 64.3 | | 1.8 | **26** | **723** |

Table 3: Classification of normal samples by both analyzing the crack lengths and the clinical diagnostic method (The analysis of the remaining carrier samples have been shown in table S4 of the ESI)

| Sl .No | Sex | HPLC Test NO. | HB | MCV | MCH | HBA2 | HBF | Crack Length (in microns) |
|---|---|---|---|---|---|---|---|---|
| 21 | Male | 71 | 14 | 81.4 | 25.7 | 3.2 | 0.2 | **916** |
| 22 | Female | 62 | 10 | 63.7 | 18.6 | 2.7 | 0.2 | **975** |
| 23 | Male | 34 | 13 | 97 | 30.7 | 3.4 | 0.2 | **840** |
| 24 | Female | 5 | 9.7 | 82.1 | 23.5 | 2.6 | 0.6 | **1032** |
| 25 | Female | 95 | 9.8 | 71.9 | 21.7 | 2.6 | 0.3 | **877** |
| 26 | Male | 83 | 13 | 68 | 20.7 | 3.5 | 3.1 | **883** |
| 27 | Male | 91 | 11 | 69.8 | 21.4 | 2.9 | 0.2 | **1013** |
| 28 | Female | 20 | 11 | 73.4 | 21.7 | 2.7 | 0.2 | **881** |
| 29 | Female | 96 | 11 | 69.3 | 20.5 | 3.2 | 0.2 | **863** |
| 30 | Female | 69 | 13 | 80 | 24 | 3 | 0.2 | **893** |
| 31 | Female | 37 | 9.7 | 76.3 | 22.8 | 3.4 | 0.3 | **861** |
| 32 | Female | 14 | 10.3 | 74.5 | 21.9 | 2.5 | 0.3 | **883** |
| 33 | Female | 81 | 11 | 72.2 | 22.2 | 3 | 0.2 | **870** |
| 34 | Male | 75 | 14 | 74.4 | 22.8 | 3.2 | 0.1 | **847** |
| 35 | Male | 59 | 12 | 82.6 | 25.3 | 2.8 | 0.2 | **911** |
| 36 | Female | 6 | 10 | 66.9 | 19.7 | 2.5 | 0.1 | **814** |
| 37 | Female | 94 | 11 | 75.8 | 22.3 | 3.3 | 0.2 | **861** |
| 38 | Male | 2 | 13.2 | 72.4 | 21.2 | 2.3 | 0.2 | **1397** |
| 39 | Female | 67 | 10 | 72.1 | 22.4 | 2.9 | 0.3 | **942** |
| 40 | Male | 51 | 14 | 84.5 | 26.3 | 3.2 | 0.3 | **916** |

Increasing the droplet volume to 4 μl prompted a proportional increase in the lengths of the radial cracks in both the carrier and the healthy samples (as shown in the figure S5 of the ESI). However, the droplet volume was optimized to 2 μl, as droplets of larger volumes subjected the crack lengths to human bias, owing to its difficulty in dispensing a perfectly circular droplet on the glass slides.

## 3.2 Determining the Threshold Crack Length for Classification of Samples

To determine the threshold crack length below which the samples can be classified as carriers, 99% confidence interval for the population of 100 samples was calculated using the following equation:

$$\text{Lower limit: } \bar{x} - 2.576\frac{\sigma}{\sqrt{n}}, \text{ Upper Limit: } \bar{x} + 2.576\frac{\sigma}{\sqrt{n}} \quad \ldots\ldots\ldots\ldots(1)$$

Where, $\bar{x}$ = average crack length of the healthy/carrier samples in microns, $\sigma$ = standard deviation and n= number of samples.

Table 4: Estimation of parameters for 99% confidence interval

| Sample | $\bar{x}$ | $\sigma$ | n | Lower Limit | Upper Limit |
|---|---|---|---|---|---|
| Healthy | 936.313 | 111.582 | 50 | 895.664 | 976.962 |
| Carrier | 652.124 | 69.839 | 50 | 626.681 | 677.566 |

(Lower Limit) $_{Healthy}$ − (Upper Limit) $_{Carrier}$ = (895.664 − 677.566) μm = 218.098 μm

Mean= 218.098/2 = 109.049 μm

(Threshold Upper Limit)$_{Carrier}$ = (677.566+109.0195) = 786.615 μm = ~800 μm

Figure 5 depicts the variation in the range of crack lengths as observed in the dried droplet patterns of the healthy and the carrier samples.

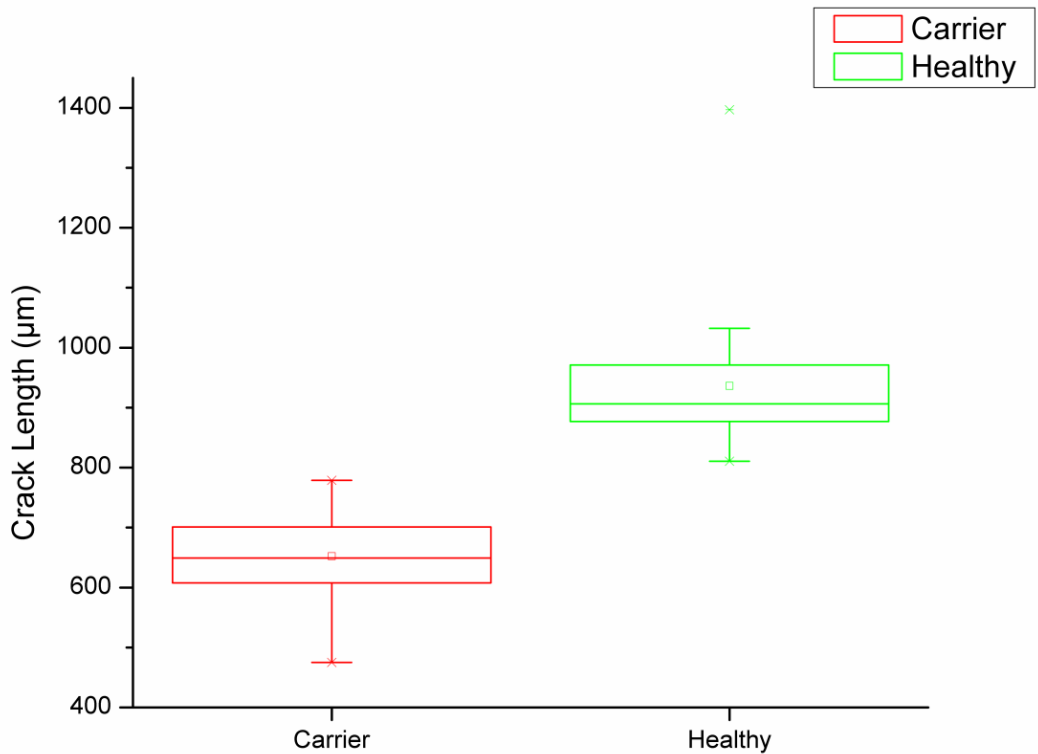

Figure 5: Variation in crack lengths in the dried drop patterns of the healthy and carrier samples

The determining crack length was estimated to be 800 μm, as shown, from the confidence interval of both the carrier and the normal samples. A sample whose average crack length was less than 800 microns was classified as a carrier sample, whereas the ones with larger radial cracks were considered as normal. Identifying the carrier samples by this image analysis technique gave zero false negative results.

### 3.3 Process Automation

In order to automate the system, a Graphical User Interface (GUI) was designed (using MATLAB R2017a) for this purpose. We named the system 'ASAP', for Automated Span Analysis of Patterns. First, the foreground droplet was separated from the noisy background by binarizing the image and detecting the longest connected boundary in the binary image. Then, the red hue from the image was removed after converting the colour palette to HSV. The image was then binarized and morphological operations of dilation[22] and closure[23] were applied using disk shaped structuring element of radii 30 and 10 pixels respectively to connect all the cracks whose links broke due to binarization and thresholding. A dilation was again performed on this image using the same structuring element as the one used for closing. This was done to avoid spurious skeletal branches in the thinned image. Then, the image was morphologically thinned[24] and branch points were detected by considering the 8 connected neighbours of each pixel. The branch points detected at the boundary of the droplet in the binary image were considered as the starting points of the cracks. The branch points were then erased from the image using a logical AND operation in a 3X3 neighbourhood of the branch point, so that each crack can be labelled as a component separately. Connected component labeling[25] was applied to the image and the components containing the start of the cracks were considered, and the corresponding end point of the labeled component was found. The Euclidean Distance between the start and the end points was taken as the length of the crack, and the pixel to length scale was taken into account while reporting the value. Values away from the median that were at least three median absolute deviations away were taken to be outliers and removed. The mean of all refined crack lengths in the image was considered for further calculations, and the sample was categorized based on this value. Figure 6 shows the steps that were followed to determine the length of the radial cracks in the dried blood drop patterns.

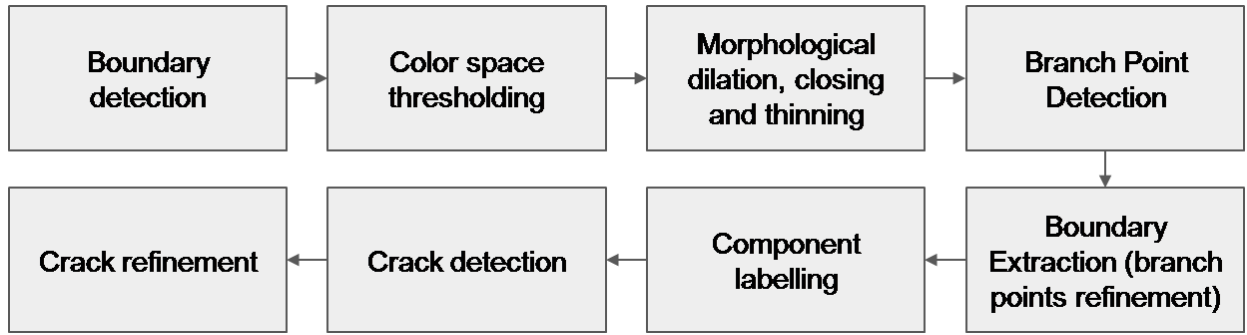

Figure 6: Steps followed for determining the crack length using using MATLAB R2017 a

'ASAP' will allow the practitioner to upload the image of the dried blood drop pattern in a computer in order to run the application. The picture can be browsed for and uploaded in the application. Once the practitioner clicks 'Evaluate', the cracks are identified, and the crack length along with an image of the detected cracks is displayed on the screen, thus, categorizing the sample as healthy or carrier. There is also a provision for analyzing a set of images at once and directly storing the images in a file instead of going through the GUI one by one for every image. Figure 7 shows the interface of the developed system 'ASAP'.

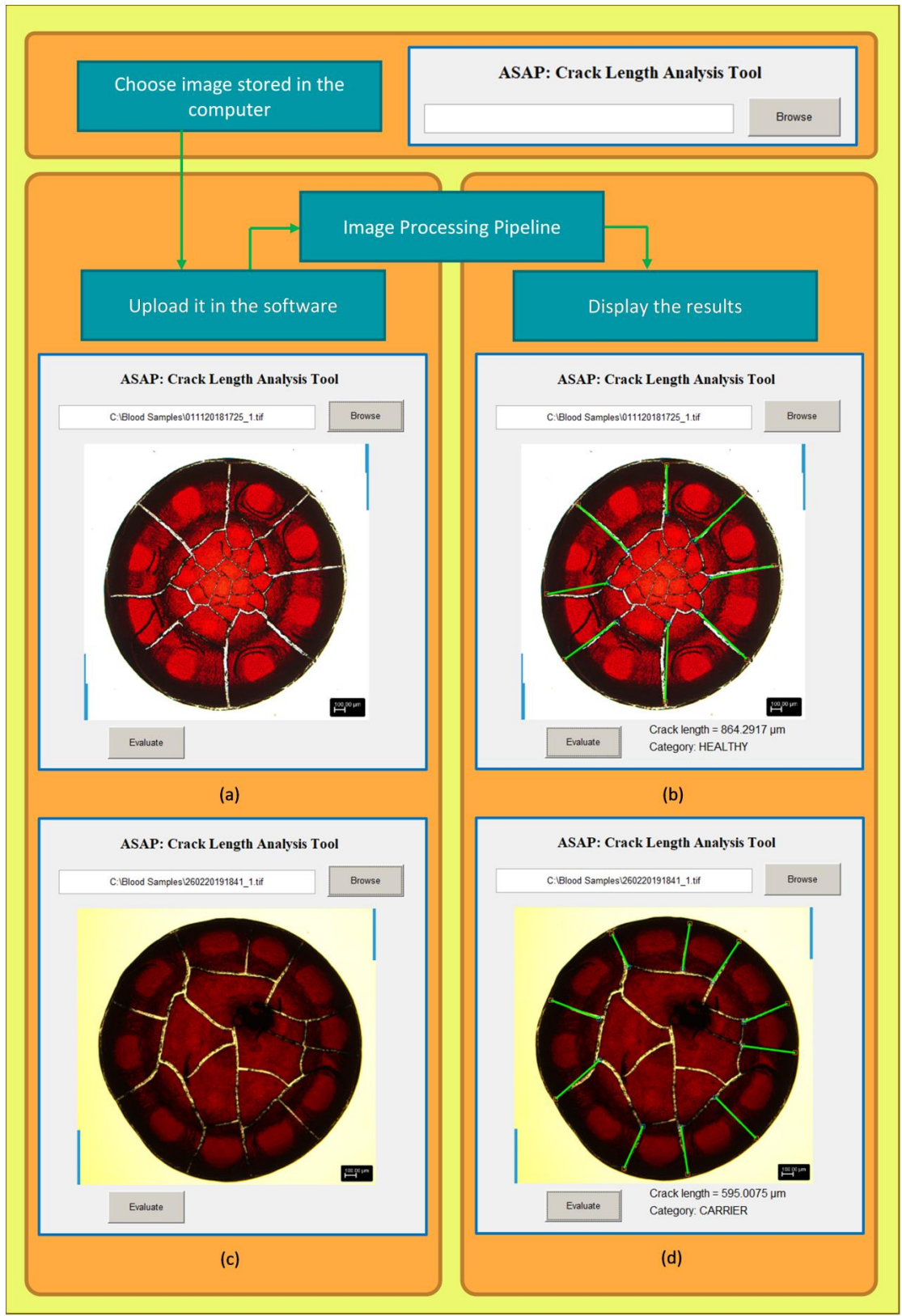

Figure 7: User interface of the developed system 'ASAP'

Table 5 depicts the comparison between the crack lengths obtained from both, the in-house developed program – ASAP and the open source software ImageJ. It is to be noted that the crack lengths determined using ASAP, resulted in a marginal error of $\pm 10\%$ when compared with the results obtained manually.

Table 5: Comparison of the crack lengths as estimated by ImageJ and ASAP

(The analysis of the remaining samples have been shown in table S6 of the ESI)

| SL.NO | Crack length as estimated by ImageJ (μm) | Crack length as detected by ASAP (μm) | Category |
|---|---|---|---|
| 1 | 677 | 712 | Carrier |
| 2 | 727 | 743 | Carrier |
| 3 | 682 | 754 | Carrier |
| 4 | 609 | 700 | Carrier |
| 5 | 683 | 700 | Carrier |
| 6 | 608 | 649 | Carrier |
| 7 | 631 | 457 | Carrier |
| 8 | 598 | 711 | Carrier |
| 9 | 618 | 769 | Carrier |
| 10 | 701 | 740 | Carrier |
| 11 | 688 | 737 | Carrier |
| 12 | 635 | 654 | Carrier |
| 13 | 643 | 665 | Carrier |
| 14 | 523 | 462 | Carrier |
| 15 | 675 | 736 | Carrier |
| 16 | 511 | 375 | Carrier |
| 17 | 627 | 313 | Carrier |
| 18 | 505 | 716 | Carrier |
| 19 | 601 | 269 | Carrier |
| 20 | 723 | 698 | Carrier |
| 21 | 916 | 914 | Healthy |
| 22 | 975 | 971 | Healthy |

| 23 | 840 | 828 | Healthy |
| 24 | 1032 | 1039 | Healthy |
| 25 | 877 | 868 | Healthy |
| 26 | 883 | 893 | Healthy |
| 27 | 1013 | 1000 | Healthy |
| 28 | 881 | 895 | Healthy |
| 29 | 863 | 847 | Healthy |
| 30 | 893 | 909 | Healthy |
| 31 | 861 | 802 | Healthy |
| 32 | 883 | 865 | Healthy |
| 33 | 870 | 889 | Healthy |
| 34 | 847 | 825 | Healthy |
| 35 | 911 | 888 | Healthy |
| 36 | 814 | 837 | Healthy |
| 37 | 861 | 838 | Healthy |
| 38 | 1397 | 1373 | Healthy |
| 39 | 942 | 917 | Healthy |
| 40 | 916 | 887 | Healthy |

**3.4 Role of interaction energies in determining the dried drop patterns**

The differences in the crack lengths as observed in the dried patterns patterns formed by the healthy and the carrier blood samples, can be attributed to the differences in the fundamental cellular level physical interactions i.e. the cell-substrate and the cell-cell interactions. Each of these interactions is majorly contributed by the repulsive double layer and the attractive van der Waals energies, which together comprise the DLVO interactions.

Considering the total number of red blood cells in a 2μl droplet (~8 * $10^6$ cells), a multilayer stacked arrangement of red blood cells has been taken into account. The cells are assumed to be spherical, in volume equivalence, and arranged in a hexagonal close packing arrangement. To facilitate adequate analysis, we have considered the interaction energies of the state of the system just before the formation of the cracks. Thus, at this state, it is reasonable to assume that major portion of the liquid in the droplet has evaporated, and the remaining liquid is present in the void spaces of the hexagonal close packing arrangement.

The cell-substrate (CS) interactions are assumed to be limited to the bottommost layer of the multilayer stacked arrangement, as the DLVO interaction energies are negligible for the cells in layers which are not in direct contact with the substrate (in this case, glass). The cell-substrate van der Waals interaction energy can be expressed as [26,27]:

$$U_{VDW\_CS} = \frac{A}{6}\left[\left[\frac{a}{h} + \frac{a}{h+2a}\right] + \ln\left(\frac{h}{h+2a}\right)\right] \dots\dots\dots (2)$$

where, A is the Hamaker constant; $h$ is the distance between the cell and substrate (as the charges between the cells and substrate are not diffused, $h$ can be considered to be in order of the Debye-Huckel length i.e. ~0.1-1nm) and $a$ is the equivalent radius of the red blood cells.

The cell-substrate electrical double layer interaction energy can be expressed as:

$$U_{EDL\_CS} = 2\pi a \left[\frac{n_\infty}{\pi r^2 h_i}\right]\frac{k_B T}{\kappa^2}\left(\varphi_1^2 + \varphi_2^2\right)\left[\frac{2\varphi_1\varphi_2}{\varphi_1^2 + \varphi_2^2}\cdot\ln\left[\frac{1+e^{-\kappa h}}{1-e^{-\kappa h}}\right] - \ln(1-e^{-2\kappa h})\right] \dots\dots\dots (3)$$

Where $n_\infty$ is the number of cells in the lowermost layer; r is the base diameter of the droplet; $h_i$ is the height of a single layer of cells ; $T$ is the temperature; $k_B$ is the Boltzmann constant; $\kappa$ is the reciprocal of Debye-Huckel length; $h$ is the distance between the cell and substrate (as the charges between the cells and substrate are not diffused, $h$ can be considered to be in order of the Debye-Huckel length i.e. ~0.1-1nm); $\varphi_1$ is the surface potential of cell and $\varphi_2$ is the surface potential of the substrate.

All the layers of cells in the system are subjected to cell-cell (CC) interactions. Due to hexagonal close packing, each cell is in contact with 12 other neighbouring cells (neglecting the edge effects). The cell-cell interactions are assumed to be solely limited to the cells in contact because the DLVO interaction energies drop to negligible values for distances between cells greater than immediate contact. The cell-cell van der Waals interaction energy can be expressed as[26]:

$$U_{VDW\_CC} = \frac{A}{6}\left[\frac{a^2}{h(2a)}\right] \quad \ldots\ldots (4)$$

Where, A is the Hamaker constant; h is the distance between two cells (considering the total length of the 55 amino acids, which are four folded, the distance can be assumed to ~10nm) and *a* is the equivalent radius of the red blood cells. The cell-cell electrical double layer interaction can be expressed as:

$$U_{EDL\_CC} = \left[12 * 2\pi a^2 \frac{k_B T}{\kappa^2}\left[\frac{2\varphi_1\varphi_2}{\varphi_1^2 + \varphi_2^2}\cdot\ln\left[\frac{1+e^{-\kappa h}}{1-e^{-\kappa h}}\right] - \ln(1-e^{-2\kappa h})\right]\left[\frac{n_\infty}{\pi r^2 h_i}\right](N_{Stack} - 2)\right]/2 \quad \ldots(5)$$

Where, *a* is the equivalent radius of the red blood cells; h is the distance between two cells (considering the total length of the 55 amino acids, which are four folded, the distance can be assumed to ~10nm); $n_\infty$ is the number of cells in a single layer; $N_{Stack}$ is the total number of stacks in the hexagonal close packed arrangement of cells; $T$ is the temperature; $k_B$ is the Boltzmann constant; $\kappa$ is the reciprocal of Debye-Huckel length, and $\varphi_1$ and $\varphi_2$ are the surface potential of erythrocytes ($\varphi_1 = \varphi_2$), r is the base diameter of the droplet and $h_i$ is the height of a single layer of cells.

The parameters that have been used for evaluation of the aforementioned total interaction energy, as used in this study, are listed in Table 6.

Table 6: Parameters used for evaluation of interaction energy

(Details regarding the estimation of parameters have been shown in S7 of ESI)

| Physical parameter | Value |
|---|---|
| Hamaker Constant (A)[28] | $7 \times 10^{-21}$ J |
| Radius of an erythrocyte ($a$) | Healthy = $2.78 \times 10^{-6}$ m <br> Carrier = $2.56 \times 10^{-6}$ m |
| Number of cells in one layer ($n_\infty$) | Healthy = 137931 <br> Carrier = 126984 |
| Surface potential of cells | Healthy = -12mV <br> Carrier = -9mV |
| Surface potential of substrate (glass) | -60mv |
| Reciprocal of Debye Huckel length ($\kappa$)[28] | $(0.78)^{-1}$ (nm)$^{-1}$ |
| Total number of stacks in the hexagonal close packed arrangement of cells ($N_{Stack}$) | Healthy = 58 <br> Carrier = 63 |
| Base diameter of the droplet (r) | $1.3 \times 10^{-3}$ m |
| Height of a single layer of cells | Healthy = $5.56 \times 10^{-6}$ m <br> Carrier = $5.12 \times 10^{-6}$ m |

The total interaction energy can be evaluated upon summing the respective van der Waals and electric double layer energy components for cell-substrate and cell-cell interactions.

Total Interaction Energy = $U_{VDW\_CS} + U_{EDL\_CS} + U_{VDW\_CC} + U_{EDL\_CC}$ ……… (6)

The interaction energies thus evaluated have been shown in Figure 8.

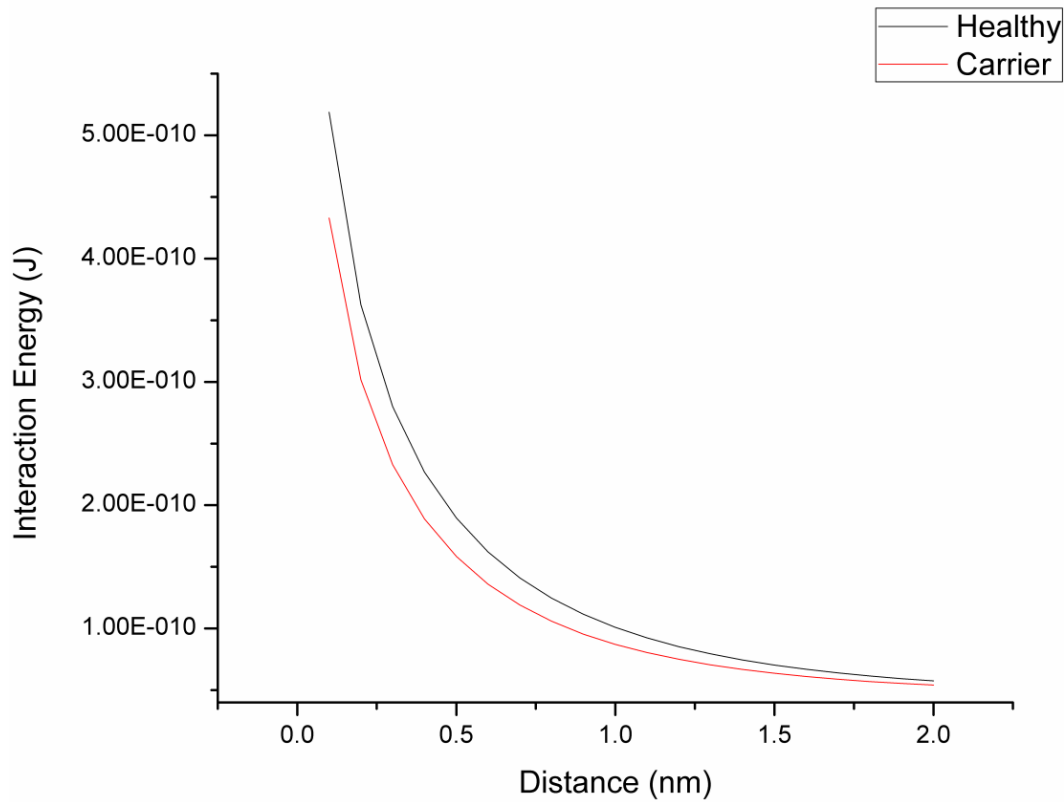

Figure 8: Variation of the total interaction energies of the healthy and the carrier samples as a function of cell-to-substrate distance

It has been discerned that the total interaction energy of the system consisting of healthy cells was distinctly higher in magnitude as compared to the carriers. This increase in the interaction energy was, hence, manifested in the form of longer cracks in the healthy samples as compared to the carriers. A direct implication of the increase in the interaction energy also resulted in early initiation of cracks in the healthy samples (as shown in Figures S8 and S9 of ESI).

## 4. CONCLUSION

The present study has conclusively proved that distinct patterns are observed on drying of whole blood droplets for carrier and normal samples. Length of the radial cracks is significantly shorter for carrier samples as compared to normal ones. This difference in the crack lengths of the dried blood droplet patterns in the healthy and the carrier samples has been explained in terms of the total interaction energy of the system, comprising the cell-substrate and the cell-cell interactions. Higher magnitude of the total interaction energy resulted in the formation of the longer cracks in the healthy samples as compared to the carriers. The in-house developed program, ASAP, automatically classified the samples as healthy or carriers based on their crack lengths. It categorized a sample as a carrier whose average crack length was less than 800 microns, whereas the ones with larger radial cracks were classified as normal. To use this technique as an on field method, these patterns can be further utilized to create a databank for automated classification of the samples, by employing appropriate techniques from computational pattern recognition and deep learning. Proposed method will examine the images of dried blood drops, extract its distinctive features and categorize as normal or carrier samples by comparing it with reference images stored in the databank. This automated process will remarkably increase the number of total population screened for thalassaemia per year in the country and will reduce the burden in the state run advanced health facilities along with a significant reduction in costs incurred for screening of each sample.

## 5. REFERENCES

1. Robins & Cotran, Pathologic Basis of Disease, *Elsevier*, 2014, Ninth edition, 638-642.

2. Antonio Cao and Renzo Galanello, *Genetics in Medicine*, 2010, 12, 61-76.

3. Doughlas R. Higgs, *Cold Spring Harb Perspect Med*, 2013, 3, 1-15.

4. Sanjay Piplani, Rahul Manan, Monika Lalit, Mridu Manjari, Tajinder Bhasin and Jasmine Bawa, *Journal of Clinical and Diagnostic Research*, 2013, 7, 2784-2787.

5. Khellil Sefiane, *Journal of Bionic Engineering*, 2010, 7, S82-S93.

6. Khellil Sefiane, *Advances in Colloid and Interface Science*, 2014, 206, 372–381.

7. Ruoyang Chen, Liyuan Zhang, Duyang Zang and Wei Shen, *Advances in Colloid and Interface Science*, 2016, 231, 1–14.

8. Luca Lanotte, Johannes Mauer, Simon Mendez, Dmitry A. Fedosov, Jean-Marc Fromental, Viviana Claveria, Franck Nicoud, Gerhard Gompper, and Manouk Abkarian, *PNAS*, 2016, 113, 13289-13294.

9. T. A. Yakhno et al., *Biophysics*, 2011, 56, 1005–1010.

10. T. A. Yakhno et al., *Technical Physics*, 2010, 55, 929–935.

11. D. Brutin, B. Sobac, B. Loquet And J. Sampol, *J. Fluid Mech*, 2011, 667, 85-95.

12. T. A. Yakhno et al., *IEEE Eng Med Bio Mag*, 2005, 24, 96-104.

13. T. A. Yakhno et al., *Technical physics*, 2009, 54, 1219-1227.

14. T. A. Yakhno et al., *J. Biomedical Science and Engineering*, 2015, 8, 1-23.

15. Curtis D. Chin, Vincent Linder and Samuel K. Sia, *Lab Chip*, 2007, 7, 41-57.

16. Xiaole Mao and Tony Jun Huang, *Lab Chip*, 2012, 12, 1412–1416.

17. Li Shen, Joshua A. Hagen and Ian Papautsky, *Lab Chip*, 2012, 12, 4240–4243.

18. Julien Reboud, Gaolian Xu, Alice Garrett, Moses Adriko, Zhugen Yang, Edridah M. Tukahebwa, Candia Rowell and Jonathan M. Cooper, *PNAS*, 2019, 116, 4834-4842.

19. Pelham A. Keahey, Mathieu L. Simeral, Kristofer J. Schroder, Meaghan M. Bond, Prince J. Mtenthaonnga, Robert H. Miros, Queen Dube and Rebecca R. Richards-Kortum, *PNAS*, 2017, 114, E10965–E10971.

20. Daniel Y. Joh, Angus M. Hucknall, Qingshan Wei, Kelly A. Mason, Margaret L. Lund, Cassio M. Fontes, Ryan T. Hill, Rebecca Blair, Zackary Zimmers, Rohan K. Achar, Derek Tseng, Raluca Gordan, Michael Freemark, Aydogan Ozcan, and Ashutosh Chilkoti, *PNAS*, 2017, 114, E7054–E7062.

21. Ashok A. Kumar, Matthew R. Patton, Jonathan W. Hennek, Si Yi Ryan Lee, Gaetana D'Alesio-Spina, Xiaoxi Yang, Julie Kanter, Sergey S. Shevkoplyas, Carlo Brugnara, and George M. Whitesides, *PNAS*, 2014, 111, 14864-14869.

22. Van den Boomgard, R and R. van Balen, *Computer Vision, Graphics, and Image Processing: Graphical Models and Image Processing*, 1992, 54, 254-258.

23. Gonzalez, R. C., R. E. Woods, and S. L. Eddins, *Digital Image Processing Using MATLAB*, Gatesmark Publishing, 2009, Second edition.

24. Lam, L., Seong-Whan Lee, and Ching Y. Suen, *IEEE Transactions on Pattern Analysis and Machine Intelligence*, 1992, 14, 869-885.

25. Haralick, Robert M., and Linda G. Shapiro, *Computer and Robot Vision, Volume I*, Addison-Wesley, 1992, 28-48.

26. M. Elimelech, J. Gregory and X. Jia, Particle Deposition and Aggregation, *Butterworth-Heinemann*, 1995, First edition, p. 299.

27. Manikuntala Mukhopadhyay, Udita Uday Ghosh, Debasish Sarkar and Sunando DasGupta, *Soft Matter*, 2018, 14, 7335-7346.

28. J. N. Israelachvili, Intermolecular ans Surface Forces, *Elsevier*, 2011, Third edition.

**SUPPORTING INFORMATION**

S1.

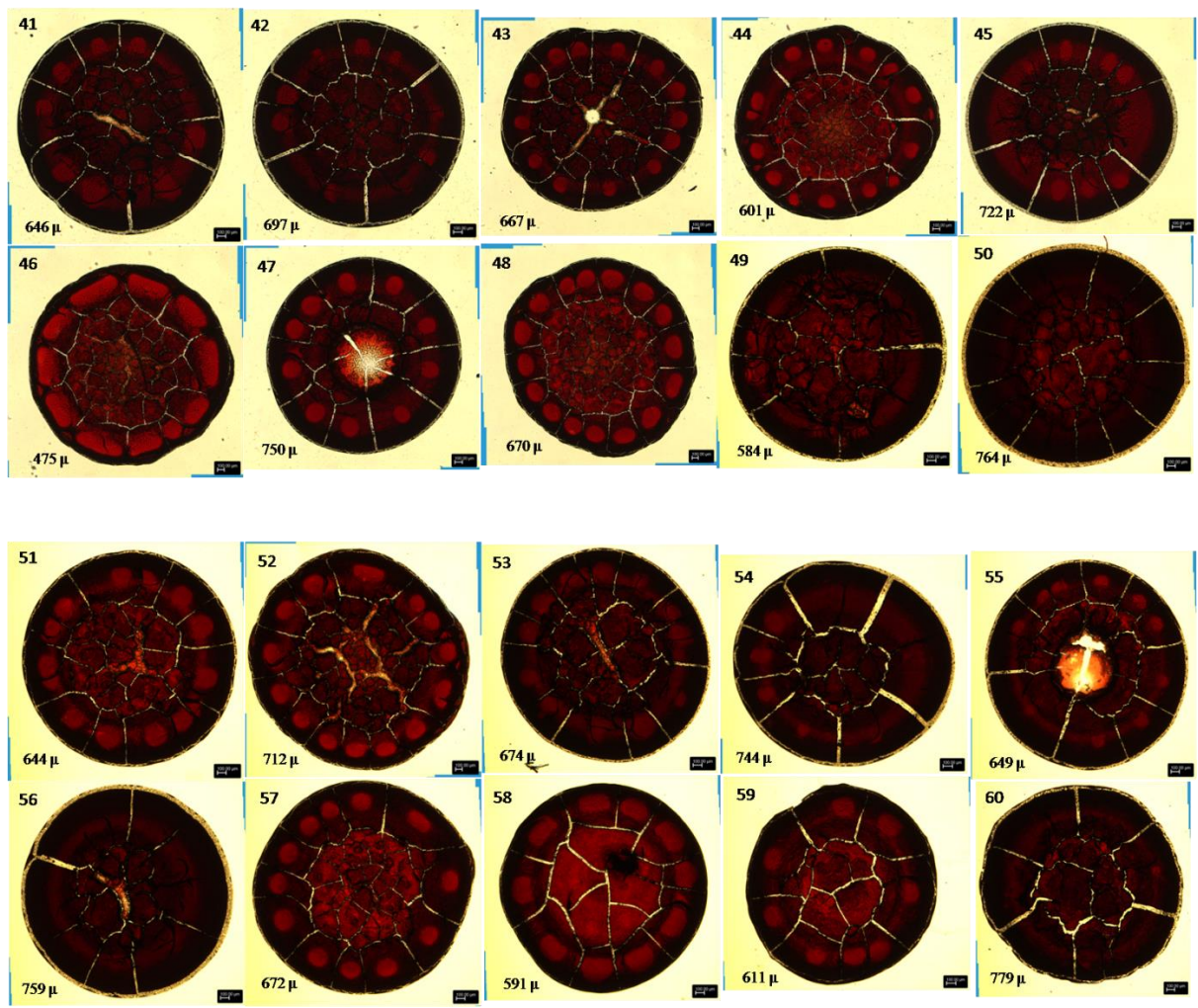

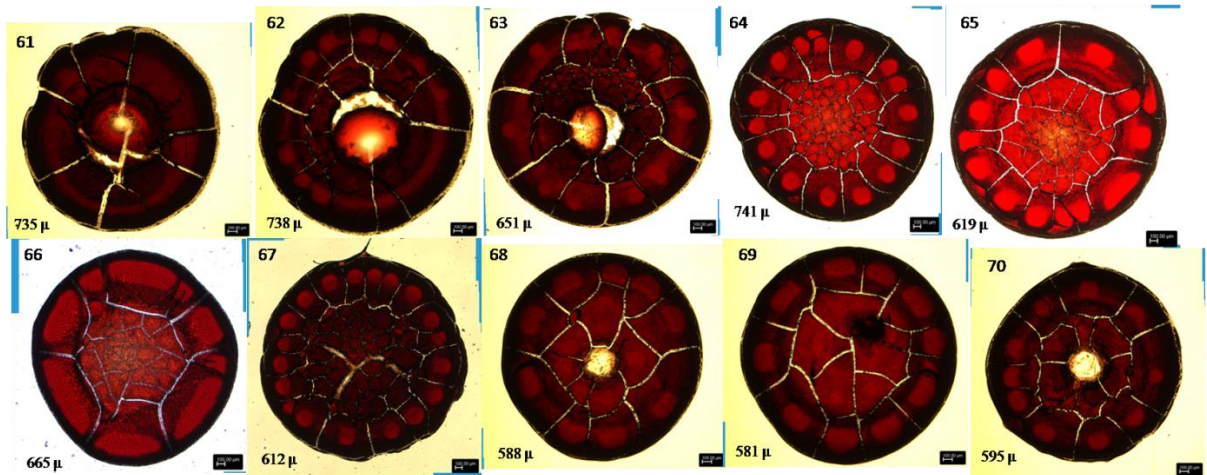

Figure S1: Dried blood drop patterns of carrier samples (the numbers on the top left denote the serial numbers whereas the ones below denote the average crack lengths)

S2.

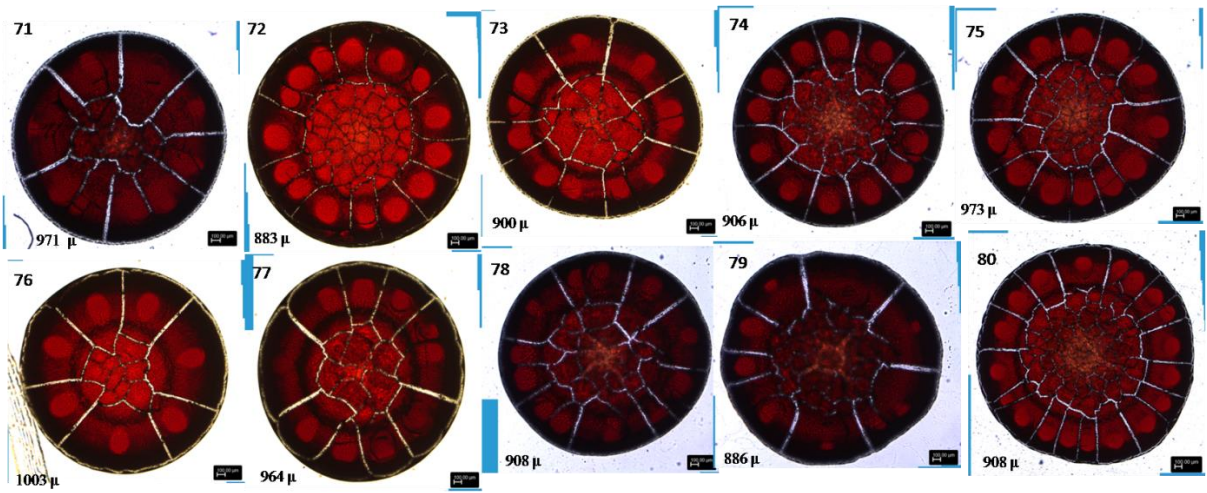

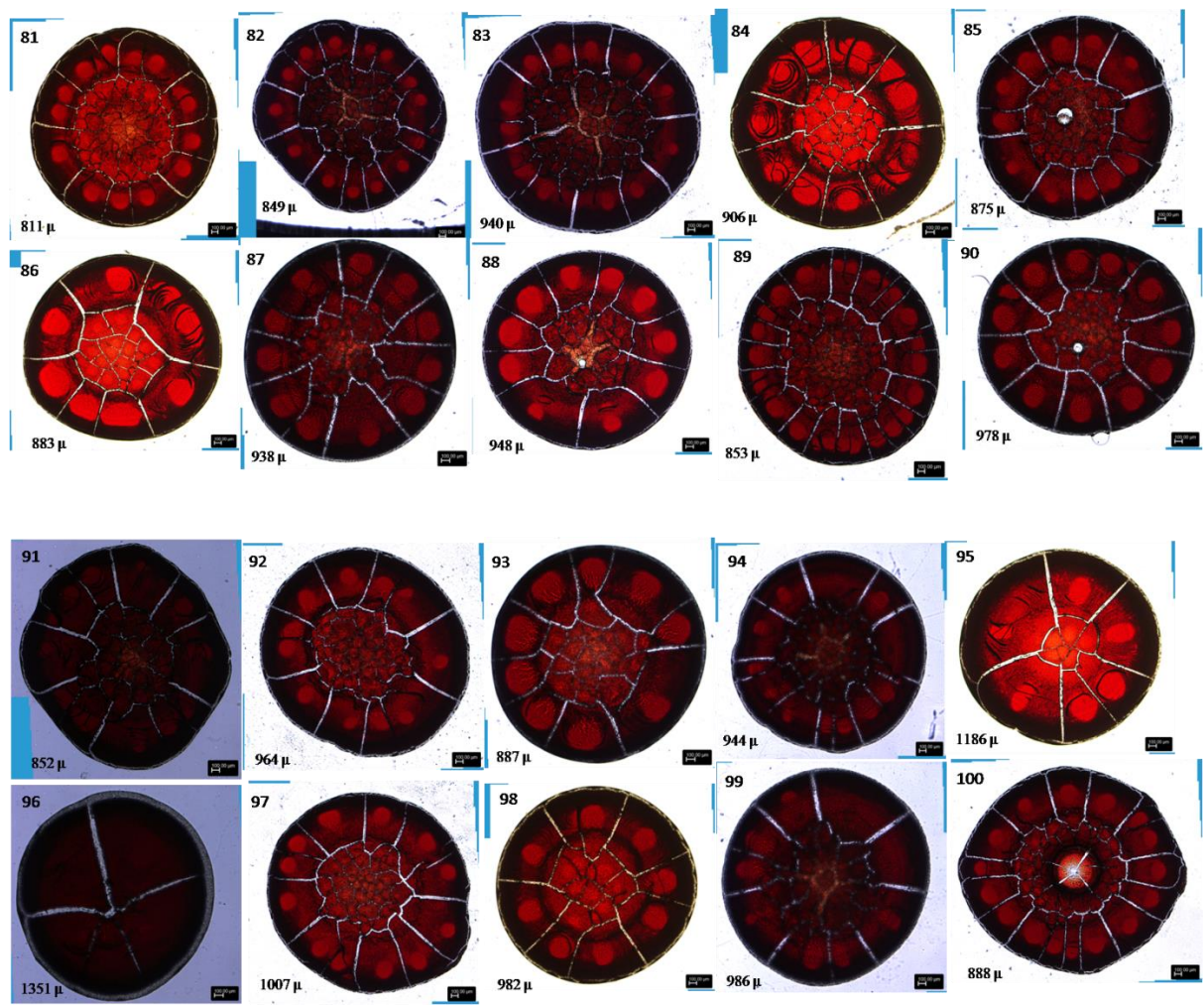

Figure S2: Dried blood drop patterns of healthy samples (the numbers on the top left denote the serial numbers whereas the ones below denote the average crack lengths)

S3. Table S3: Classification of carrier samples by both analyzing the crack lengths and the clinical diagnostic method

| SL.No | Age | Sex | **Diagnosis** | HPLC Test No. | HB | MCV | MCH | HBA0 | **HBA2** | HBF | **HBA2e** | **Crack length (in microns)** |
|---|---|---|---|---|---|---|---|---|---|---|---|---|
| 41 | 30 | Male | **Beta thalassaemia carrier** | 320 | 13.1 | 64.8 | 20 | 85.2 | **5.7** | 0.8 | | **646** |
| 42 | 37 | Male | **Beta thalassaemia carrier** | 340 | 12.7 | 68 | 20.9 | 83.9 | **5.2** | 0.4 | | **697** |
| 43 | 20 | Male | **HbE carrier** | 354 | 13.3 | 74.2 | 24.2 | 59.5 | | 0.8 | **30.4** | **667** |
| 44 | 27 | Male | **Beta thalassaemia** | 368 | 12 | 63.8 | 18.8 | 85.9 | **5.4** | 0.3 | | **601** |
| | | | carrier | | | | | | | | | |
| 45 | 31 | Male | **Beta thalassaemia carrier** | 526 | 13.6 | 63.9 | 19.2 | 85.7 | **5.1** | 0.4 | | **722** |
| 46 | 20 | Female | **HbE carrier** | 533 | 9.6 | 75.2 | 22.1 | 66.4 | | 0.8 | **25.5** | **475** |
| 47 | 26 | Male | **Beta thalassaemia carrier** | 548 | 13.4 | 67 | 20.4 | 85.1 | **5.1** | 0.3 | | **750** |
| 48 | 32 | Female | **HbE carrier** | 572 | 9.9 | 73.1 | 23.6 | 64.2 | | 0.6 | **27.1** | **670** |
| 49 | 25 | Female | **HbE carrier** | 2 | 12.1 | 76.2 | 25.9 | 63.8 | | 0.5 | **26.3** | **584** |
| 50 | 34 | Male | **Beta thalassaemia carrier** | 14 | 11.8 | 64.8 | 20.1 | 79.5 | **5** | 0.8 | | **764** |
| 51 | 28 | Female | **HbE carrier** | 28 | 11.1 | 80.7 | 26.2 | 61.8 | | 0.3 | **21.3** | **644** |
| 52 | 15 | Female | **Beta thalassaemia carrier** | 32 | 10.9 | 82.4 | 25.2 | 86.6 | **5** | 0.2 | | **712** |
| 53 | 25 | Female | **HbE carrier** | 44 | 11.3 | 82.6 | 28 | 56.9 | | 0.3 | **25.5** | **674** |
| 54 | 30 | Male | **HbE carrier** | 53 | 14.4 | 79.2 | 26.8 | 63.6 | | 0.7 | **25.5** | **744** |
| 55 | 24 | Female | **HbE carrier** | 59 | 11.3 | 76.4 | 24.9 | 64.7 | | 0.6 | **24.4** | **649** |
| 56 | 24 | Male | **Beta thalassaemia carrier** | 75 | 14 | 69.4 | 21.3 | 85 | **4.8** | 0.4 | | **759** |
| 57 | 20 | Female | **HbE carrier** | 82 | 11.6 | 80.2 | 27 | 65.4 | | 0.8 | **24.4** | **672** |
| 58 | 23 | Female | **Beta thalassaemia carrier** | 102 | 8.3 | 69.1 | 21.4 | 86.3 | **4.6** | 0.7 | | **591** |
| 59 | 19 | Female | **HbE carrier** | 135 | 9.8 | 78.5 | 25.7 | 68.3 | | 0.6 | **25.1** | **611** |
| 60 | 24 | Female | **HbE carrier** | 147 | 11.2 | 75.9 | 24.8 | 66 | | 0.8 | **25.5** | **779** |
| 61 | 26 | Male | **HbE carrier** | 165 | 13.5 | 76.4 | 24.9 | 65.2 | | 0.5 | **27.4** | **735** |
| 62 | 19 | Female | **HbE carrier** | 172 | 11.4 | 74.2 | 25.4 | 66 | | 0.6 | **27.3** | **738** |
| 63 | 24 | Male | **HbE carrier** | 175 | 13.8 | 73.8 | 24.9 | 66 | | 0.9 | **26.4** | **651** |
| 64 | 16 | Female | **Beta thalassaemia carrier** | 28 | 11 | 75.6 | 22.4 | | **5.3** | 0.5 | | **741** |
| 65 | 18 | Male | **Beta thalassaemia carrier** | 44 | 12 | 67.1 | 19.7 | | **5.4** | 0.2 | | **619** |
| 66 | 16 | Female | **Beta thalassaemia carrier** | 61 | 7.7 | 61.9 | 17.7 | | **5.2** | 0.3 | | **665** |
| 67 | 16 | Female | **Beta thalassaemia carrier** | 13 | 11 | 75.3 | 22.7 | | **4.7** | 0.4 | | **612** |
| 68 | 18 | Female | **Beta thalassaemia carrier** | 47 | 11 | 77.6 | 23.2 | | **5.4** | 1.1 | | **588** |
| 69 | 12 | Male | **Beta thalassaemia carrier** | 89 | 11 | 69 | 1.2 | | **5.3** | 0.2 | | **581** |
| 70 | 24 | Male | **HbE carrier** | 124 | 11.9 | 76.5 | 25.2 | 64.2 | | 0.7 | **28** | **595** |

S4. Table S4: Classification of healthy samples by both analyzing the crack lengths and the clinical diagnostic method

| Sl .No | Sex | HPLC Test NO. | HB | MCV | MCH | HBA2 | HBF | Crack Length (in microns) |
|---|---|---|---|---|---|---|---|---|
| 71 | Male | 70 | 15 | 79 | 26 | 3 | 0.2 | 971 |
| 72 | Male | 84 | 12 | 86.3 | 27.1 | 3.3 | 0.2 | 883 |
| 73 | Male | 72 | 12 | 70.8 | 21.9 | 2.8 | 0.2 | 900 |
| 74 | Female | 64 | 11 | 73 | 22.4 | 3 | 0.2 | 906 |
| 75 | Male | 57 | 12 | 75.5 | 23.3 | 3.3 | 0.3 | 973 |
| 76 | Male | 74 | 15 | 82.6 | 26.4 | 2.8 | 0.2 | 1003 |
| 77 | Male | 73 | 14 | 74.1 | 22.7 | 2.9 | 0.1 | 964 |
| 78 | Female | 7 | 11.1 | 82.5 | 25.3 | 2.6 | 0.2 | 908 |
| 79 | Male | 22 | 12 | 77.8 | 23.8 | 2.6 | 0.2 | 886 |
| 80 | Male | 56 | 12 | 76.1 | 23.3 | 3 | 0.1 | 908 |
| 81 | Male | 87 | 13 | 70.1 | 21.9 | 2.5 | 0.2 | 811 |
| 82 | Male | 17 | 12 | 85.3 | 25.5 | 2.5 | 0.2 | 849 |
| 83 | Male | 19 | 13 | 78.7 | 23.7 | 2.7 | 0.2 | 940 |
| 84 | Male | 93 | 12 | 67.1 | 19.9 | 3 | 0.3 | 906 |
| 85 | Male | 60 | 13 | 68.3 | 20.8 | 2.9 | 0.2 | 875 |
| 86 | Female | 100 | 11 | 69.6 | 21.7 | 2.9 | 0.1 | 883 |
| 87 | Female | 66 | 12 | 72.3 | 21.7 | 3 | 0.3 | 938 |
| 88 | Female | 36 | 11 | 72.2 | 20.9 | 2.9 | 0.3 | 948 |
| 89 | Male | 58 | 12 | 67.1 | 20.6 | 3.2 | 0.2 | 853 |
| 90 | Female | 63 | 12 | 81.7 | 26.5 | 3 | 0.3 | 978 |
| 91 | Male | 1 | 12.7 | 84.1 | 25.8 | 2.4 | 0.1 | 852 |
| 92 | Male | 55 | 12 | 69.1 | 21.1 | 3.3 | 0.7 | 964 |
| 93 | Female | 65 | 9.5 | 64.8 | 17.9 | 2.8 | 0.2 | 887 |
| 94 | Male | 25 | 14 | 72.3 | 21.1 | 2.8 | 0.4 | 944 |
| 95 | Female | 99 | 12 | 70.1 | 21.1 | 2.9 | 0.2 | 1186 |
| 96 | Male | 3 | 13.7 | 72.9 | 21.7 | 2.8 | 0.3 | 1351 |
| 97 | Female | 54 | 13 | 84.1 | 27.2 | 2.8 | 0.2 | 1007 |
| 98 | Male | 76 | 13 | 85.5 | 27.9 | 3.3 | 0.4 | 982 |
| 99 | Male | 29 | 13 | 76.4 | 23.4 | 2.6 | 0.2 | 986 |
| 100 | Male | 52 | 14 | 82.8 | 25.7 | 2.6 | 0.2 | 888 |

S5.

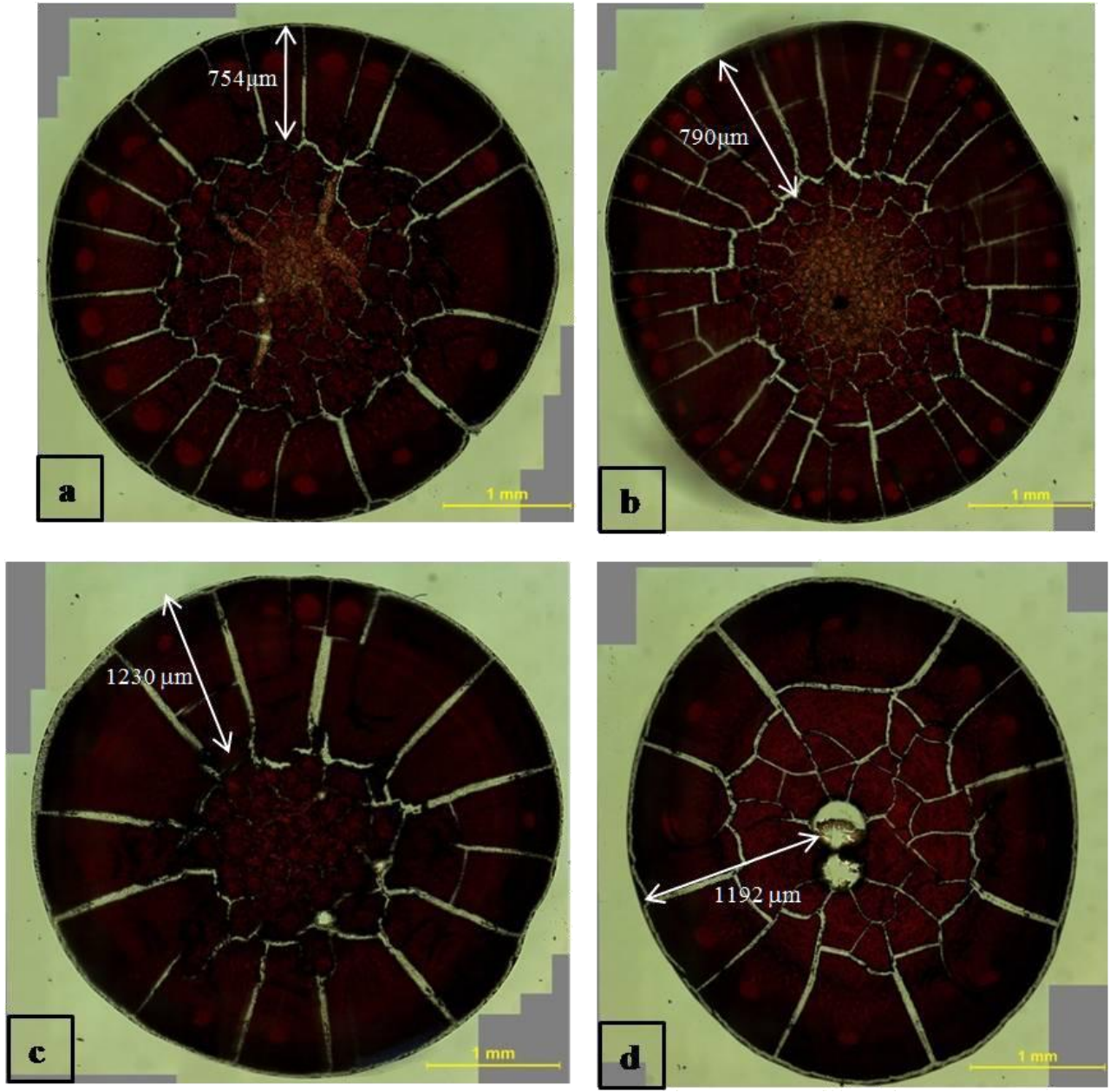

**Figure S5: Drop drying pattern of whole blood (4 μl) on glass substrates** [(a) & (b) carrier samples; (c) & (d) healthy samples ]

S6. Table S6: Comparison of the crack lengths as estimated by ImageJ and ASAP

| SL.NO | Crack length as estimated by ImageJ (μm) | Crack length as detected by ASAP (μm) | Category |
|---|---|---|---|
| 41 | 646 | 620 | Carrier |
| 42 | 697 | 756 | Carrier |
| 43 | 667 | 712 | Carrier |
| 44 | 601 | 642 | Carrier |
| 45 | 722 | 760 | Carrier |
| 46 | 475 | 62 | Carrier |
| 47 | 750 | 754 | Carrier |
| 48 | 670 | 667 | Carrier |
| 49 | 584 | 715 | Carrier |
| 50 | 764 | 777 | Carrier |
| 51 | 644 | 703 | Carrier |
| 52 | 712 | 708 | Carrier |
| 53 | 674 | 744 | Carrier |
| 54 | 744 | 722 | Carrier |
| 55 | 649 | 607 | Carrier |
| 56 | 759 | 773 | Carrier |
| 57 | 672 | 722 | Carrier |
| 58 | 591 | 619 | Carrier |
| 59 | 611 | 665 | Carrier |
| 60 | 779 | 707 | Carrier |
| 61 | 735 | 739 | Carrier |
| 62 | 738 | 752 | Carrier |
| 63 | 651 | 776 | Carrier |
| 64 | 741 | 703 | Carrier |
| 65 | 619 | 638 | Carrier |
| 66 | 665 | 779 | Carrier |
| 67 | 612 | 740 | Carrier |
| 68 | 588 | 618 | Carrier |
| 69 | 581 | 618 | Carrier |
| 70 | 595 | 678 | Carrier |
| 71 | 971 | 936 | Healthy |
| 72 | 883 | 919 | Healthy |
| 73 | 900 | 864 | Healthy |
| 74 | 906 | 868 | Healthy |
| 75 | 973 | 935 | Healthy |
| 76 | 1003 | 965 | Healthy |
| 77 | 964 | 1006 | Healthy |
| 78 | 908 | 866 | Healthy |
| 79 | 886 | 844 | Healthy |
| 80 | 908 | 864 | Healthy |
| 81 | 811 | 856 | Healthy |
| 82 | 849 | 802 | Healthy |
| 83 | 940 | 892 | Healthy |
| 84 | 906 | 957 | Healthy |
| 85 | 875 | 823 | Healthy |
| 86 | 883 | 830 | Healthy |
| 87 | 938 | 874 | Healthy |
| 88 | 948 | 928 | Healthy |
| 89 | 853 | 781 | Healthy |
| 90 | 978 | 903 | Healthy |
| 91 | 852 | 927 | Healthy |
| 92 | 964 | 887 | Healthy |
| 93 | 887 | 809 | Healthy |
| 94 | 944 | 863 | Healthy |
| 95 | 1186 | 1102 | Healthy |
| 96 | 1351 | 1266 | Healthy |
| 97 | 1007 | 916 | Healthy |
| 98 | 982 | 887 | Healthy |
| 99 | 986 | 886 | Healthy |
| 100 | 888 | 935 | Healthy |

S7. The height of a dried drop of whole blood was estimated by a surface profilometer (3D non contact surface profilometer, Contour- GTK), Bruker) and was found to be ~ 320 microns. The image of a dried blood droplet has been shown in Figure S7.

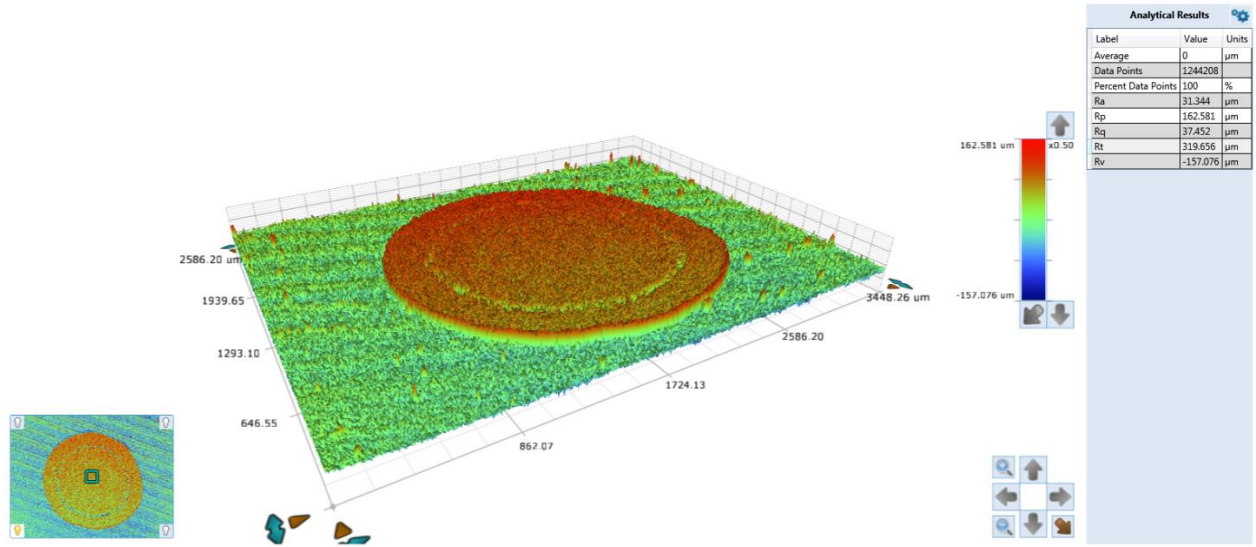

Figure S7: Image of a dried blood droplet

For healthy samples,

Total number of stacks in the hexagonal close packed arrangement of cells, $N_{Stack}$

= (Height of dried blood drop) / (Diameter of each cell)

= $(320 \times 10^{-6})/(2.78 \times 10^{-6} \times 2)$

= 58

Total number of cells in a 2µl droplet = $8 \times 10^6$ cells

So, number of cells in a single layer ($n_\infty$) = $(8 \times 10^6/58)$ = 137931

Similarly, for carrier samples, $N_{Stack}$ = $(320 \times 10^{-6})/(2.56 \times 10^{-6} \times 2)$ = 63

$n_\infty$ = $(8 \times 10^6/63)$ = 126984

S8.

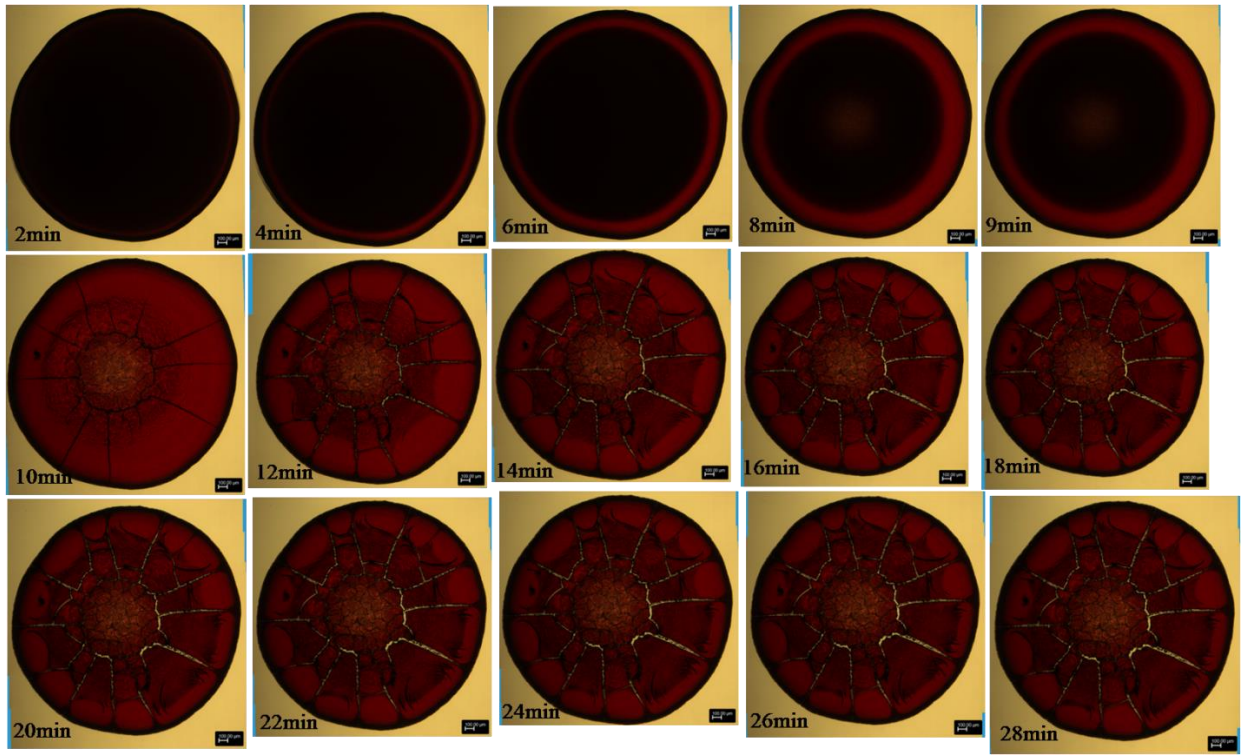

Figure S8: Crack formation in a carrier sample

S9.

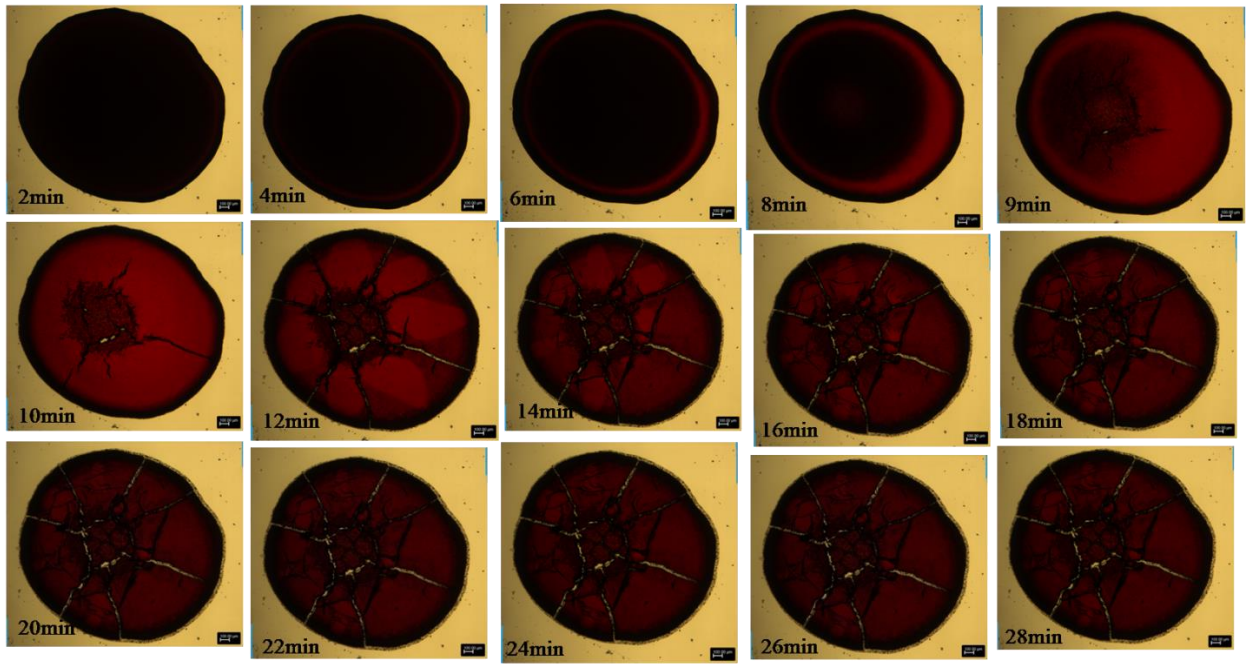

Figure S8: Crack formation in a healthy sample